\documentclass[a4paper]{jpconf}
\usepackage{graphicx}
\usepackage{iopams}
\begin{document}
\title{A multiprecision C++ library for matrix-product-state
simulation of quantum computing: Evaluation of numerical errors}
\author{Akira SaiToh}
\address{Quantum Information Science Theory Group,
National Institute of Informatics, 2-1-2 Hitotsubashi,
Chiyoda, Tokyo 101-8430, Japan}
\ead{akirasaitoh@nii.ac.jp}

\begin{abstract}
The time-dependent matrix-product-state (TDMPS) simulation method has been
used for numerically simulating quantum computing for a decade. We introduce
our C++ library ZKCM\_QC developed for multiprecision TDMPS simulations of quantum
circuits.
Besides its practical usability, the library is useful for evaluation of the
method itself. With the library, we can capture two types of numerical errors
in the TDMPS simulations: one due to rounding errors caused by the shortage in
mantissa portions of floating-point numbers; the other due to truncations of
nonnegligible Schmidt coefficients and their corresponding Schmidt vectors. We
numerically analyze these errors in TDMPS simulations of quantum computing.
\end{abstract}

\section{Introduction}\label{secintro}
Simulations of time-evolving quantum states using a matrix-product-state
(MPS) \cite{WH93} representation have been widely used in a variety of physical
systems \cite{Review1,Sc11}. We have been developing a C++ library, named
ZKCM\_QC \cite{ZKCMQC}, for multiprecision time-dependent MPS (TDMPS)
simulations of quantum computing using Vidal's representation \cite{V03}
for MPS.
Here, we define the precision by the length of a mantissa
portion for each floating point number.

Let us begin with a brief description of conventions.
We employ the computational basis $\{|0\rangle, |1\rangle\}^n$
for $n$-qubit quantum states, where
$|0\rangle=\left(\begin{array}{cc}1 & 0\end{array}\right)^{\rm T}$ and
$|1\rangle=\left(\begin{array}{cc}0 & 1\end{array}\right)^{\rm T}$.
An $n$-qubit quantum state is represented as
$
|\Psi\rangle=\sum_{i_0\cdots i_{n-1}=0\cdots0}^{1\cdots1}
c_{i_0\cdots i_{n-1}}|i_0\cdots i_{n-1}\rangle
$
with complex amplitudes $c_{i_0\cdots i_{n-1}}$. In this paper,
we employ the following MPS form \cite{V03,SK06} of the state, for our
TDMPS simulations.
\begin{equation}
\begin{array}{rl}\label{eqMPS}
 |\Psi\rangle=&\sum_{i_0\cdots i_{n-1}=0\cdots0}^{1\cdots1}\biggl[
\sum_{v_0=0}^{m_0-1}\sum_{v_1=0}^{m_1-1}\cdots\sum_{v_{n-2}=0}^{m_{n-2}-1}
Q_0(i_0,v_0)V_0(v_0)Q_1(i_1,v_0,v_1)V_1(v_1)\cdots \\ &
Q_{s}(i_{s},v_{s-1},v_s)V_s(v_s)\cdots V_{n-2}(v_{n-2})Q_{n-1}(i_{n-1},v_{
n-2})\biggr]
|i_0\cdots i_{n-1}\rangle,
\end{array}
\end{equation}
where we use tensors $\{Q_s\}_{s=0}^{n-1}$ with parameters $i_s, v_{s-1}, v_s$
($v_{-1}$ and $v_{n-1}$ are excluded) and $\{V_s\}_{s=0}^{n-2}$ with parameter
$v_s$; $m_s$ is a suitable number of values for $v_s$,
with which the state is represented precisely or well approximated.
[Tensor $V_s(v_s)$ stores the Schmidt coefficients for the
splitting between the $s$th site and the $(s+1)$th site.]

With the MPS form, the cost to update data in accordance with a time
evolution is reduced considerably in comparison to the brute-force method.
We have only to update tensors corresponding to the sites under the
influence of a unitary operator for each step.
The cost to simulate an evolution by a single unitary operation
$\in {\rm U}(4)$ acting on some consecutive sites is
$O(m_{\rm max}^3)$ floating-point operations where $m_{\rm max}$ is
the largest value of $m_s$ among the sites $s$ \cite{V03}.
The total cost of a TDMPS simulation thus grows polynomially in the
largest Schmidt rank among those for the splittings, which highly
depends on the instance of the problem of one's concern.
(Here, each splitting we consider is, of course, a bipartite splitting
between consecutive sites.)

The TDMPS simulation method in double precision [{\em i.e.}, (52+1)-bits-long
mantissa for a floating point number] has already been used commonly in the
community of computational physics \cite{ALPS}. There are, however, known
cases for general computational methods where more accurate computation is
needed to obtain a reliable results describing physical phenomena \cite{BBB12}.
Accumulations of rounding errors of basic arithmetic operations are the
main cause in such cases. This should be true also in TDMPS simulations of
quantum computing where many matrix diagonalizations are involved unless the
depth of quantum circuits is very small.
Besides the precision of basic operations, another factor of losing accuracy
is the truncation of Schmidt coefficients. When many of Schmidt coefficients are
nonnegligible at a certain step of a TDMPS simulation, imposing a threshold to the number
of Schmidt coefficients for each splitting may truncate out important data
affecting simulation results \cite{VDMB09}. So far truncations of nonvanishing
Schmidt coefficients have been uncommon\footnote{
We are discussing on the standard circuit model of quantum computing. It was reported
\cite{Banuls06} that truncations of nonzero Schmidt coefficients were employed in a
TDMPS simulation in a Hamiltonian model for adiabatic quantum computing.
} and the largest Schmidt rank and its upper
bound in the absence of truncations have been of main concern when MPS and related
data structures are used for handling quantum and/or classical computational
problems \cite{V03,KW04,MS08,SK06,YS06,J06,CM12,TW12,JBC12}.
This is in contrast to MPS and TDMPS simulations in condensed matter physics where
truncations are very commonly employed \cite{H06}.

In this report, we evaluate numerical errors in actual TDMPS simulations
of quantum computing. We first begin with a brief introduction to the
ZKCM\_QC library in section \ref{seczkcmqc}. Then we conduct numerical
evaluations in section \ref{secne}: an error due to precision
shortage is investigated in section\ \ref{secprec} and that due to
truncations of Schmidt coefficients is investigated in section\ \ref{sectrunc}.
We discuss and summarize obtained results in sections\ \ref{secdiscussion}
and \ref{secconclusion}, respectively.

\section{ZKCM\_QC library}\label{seczkcmqc}
The ZKCM\_QC library has been developed as a C++ library for TDMPS simulations of
quantum computing with an emphasis on multiprecision computation. It is an extension
package of the ZKCM library \cite{ZKCM} which uses the GMP \cite{GMP} and MPFR \cite{MPFR}
libraries for basic arithmetic operations. We briefly explain the usage of the ZKCM\_QC
library.

The library can be installed by the standard process: ``\verb|./configure|'', ``\verb|make|'',
and ``\verb|make install|'' in any Unix-like system with GNU tools. Once it is installed,
a user will write a C++ program using the header file ``\verb|zkcm_qc.hpp|'' and
compile the program with the library flags ``\verb|-lzkcm_qc -lzkcm -lmpfr -lgmp -lgmpxx -lm|''
using a C++ compiler.

In the main part of a program, a user will firstly specify the default precision.
For example ``\verb|zkcm_set_default_prec(256);|'' will set the default precision to 256 bits.
Then, a user will call a constructor of the class
``\verb|mps|'' to create the object of an MPS kept in the form of (\ref{eqMPS}). For example,
``\verb|mps M(5);|'' will generate an object keeping the data of $|0_00_10_20_30_4\rangle$
in the MPS form. Then a user will apply certain unitary operations by using some member
functions of the class, such as ``\verb|applyU|'' and ``\verb|applyU8|'' with
predefined and/or user-defined unitary matrices. For example,
``\verb|M.applyU(tensor2tools::Hadamard, 2);|'' applies an Hadamard gate $H =
\left(\begin{array}{cc}1 & 1\\ 1 & -1\end{array}\right)/\sqrt{2}$ to qubit $2$.
For another example, ``\verb|M.applyU(tensor2tools::CNOT, 2, 4);|'' applies the CNOT gate
to qubits $2$ and $4$ that are the control bit and the target bit, respectively.
Here, CNOT is a conditional bit flip; the target bit is flipped when the control bit is
in $|1\rangle$. There are other member functions for unitary operations. For instance,
``\verb|CCNOT|'' is used for a CCNOT operation; ``\verb|M.CCNOT(0, 2, 4);|'' applies
a bit flip on qubit $4$ under the condition that qubits $0$ and $2$ are both in $|1\rangle$.
In addition, it is possible to simulate a projective measurement by calling the member function
``\verb|pmeasure|''. For example, ``\verb|int event = M.pmeasure(zkcm_matrix("[1,0;0,-1]"), 2, -1);|''
will simulate a projective measurement on qubit $2$ with the observable Pauli $Z$, as a random
process. The returned value ``\verb|event|'' will be $0$ when the event corresponding to the
larger eigenvalue of $Z$ occurs and will be $1$ otherwise.
There are other useful functions documented in the reference manual of the library. Among them,
functions to achieve a reduced density matrix of specified qubits will be frequently used.
One example to use such a function is ``\verb|std::string s = M.RDO_block(2, 4).str_dirac_b();|''
which obtains the reduced density matrix of the block of qubits $2$, $3$, and $4$ as a
``\verb|std::string|''-type string.

For each call of unitary operations and/or projective measurements, we update
involved tensors of the MPS. This is a tedious process, but is concealed by the library.
User programs usually do not pay attention to the background simulation process.
One may still set a threshold $m_{\rm trunc}$ on the number of Schmidt coefficients of a
bipartite splitting at one's risk. A user will write, {\em e.g.}, ``\verb|M.m_trunc(12);|'' to
specify the threshold (it is set to $12$ for this example, {\em i.e.}, only largest $12$ Schmidt
coefficients will be kept at each time tensors are updated during the TDMPS simulation).

More detailed instructions on the installation and the usage of the ZKCM\_QC library are found in the
documents placed at the ``\verb|doc|'' directory of the package.

\section{Numerical errors in TDMPS simulations of quantum computing}\label{secne}
In this section, we firstly investigate the influence of precision shortage
in TDMPS simulations of a simple quantum search \cite{Grover96} and a quantum
Fourier transform \cite{Shor}; we secondly investigate the influence of the truncation
of Schmidt coefficients in a TDMPS simulation of the Deutsch-Jozsa algorithm \cite{DJ92}
for a simple boolean function.
\subsection{Errors due to the precision shortage}\label{secprec}
A TDMPS simulation is in fact sensitive to the precision of floating-point
computation. Here, we show typical examples.

\subsubsection{Example 1}\label{secprec1}
\begin{figure}[hbt]
\begin{center}
\includegraphics[width=95mm]{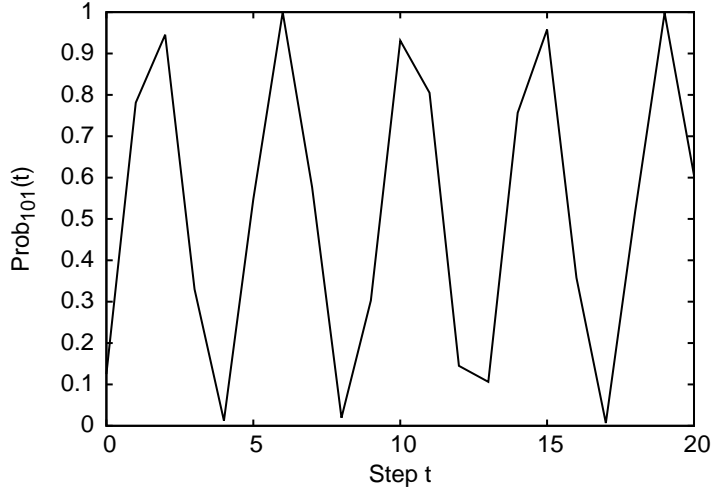}
\caption{\label{figosc}Theoretical graph of the oscillation of
${\rm Prob}_{101}(t)$ in the three-qubit quantum search described in the text.}
\end{center}
\end{figure}

\begin{figure}[hbt]
\begin{center}
\includegraphics[width=105mm]{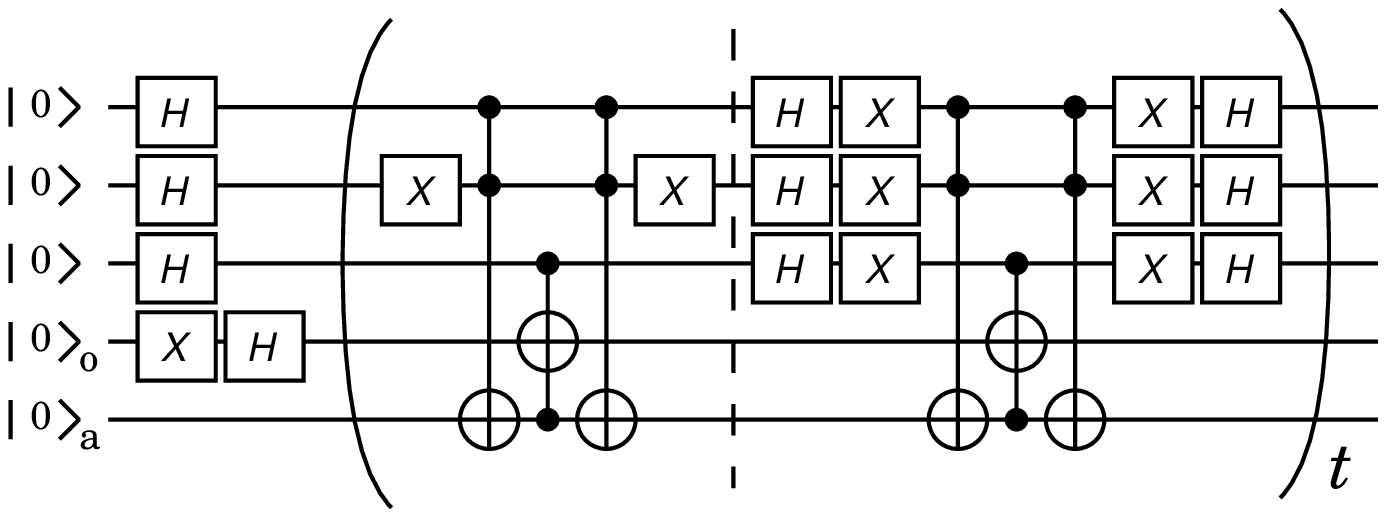}
\caption{\label{figGC}Quantum circuit of the three-qubit quantum search described
in the text. The circuit in the parentheses illustrates $R$, in which the left half
corresponds to $(1-2|101\rangle\langle101|)$ and the right half corresponds to
$(1-2|s\rangle\langle s|)$. Symbol ``o'' stands for the oracle qubit and ``a'' stands
for the ancilla qubit.}
\end{center}
\end{figure}

For the first example, we consider Grover's quantum search \cite{Grover96} (or quantum
amplitude amplification) for a simple oracle\footnote{
A TDMPS simulation of the Grover search for a simple oracle was first performed
by Kawaguchi {\em et al.} \cite{KW04} in 2004.
} with a three-qubit input.
The initial state is set to $|s\rangle = \frac{1}{\sqrt{8}}\sum_{x_0x_1x_2=000}^{111}|x_0x_1x_2\rangle$
and the target state is set to $|101\rangle$. We successively apply the so-called Grover routine\footnote{
It is in general $(1-2|s\rangle\langle s|)(1-2\sum_{x\in X}|x\rangle\langle x|)$ with $|s\rangle$ the
equally-weighted superposition of states in a parent set and $X$ the set of target states. It is highly
dependent on a problem instance how this operation is constructed as a quantum circuit.}
$R=(1-2|s\rangle\langle s|)(1-2|101\rangle\langle101|)$ to the state starting from $|s\rangle$.
This results in an oscillation of the population of $|101\rangle$,
${\rm Prob}_{101}(t)=|\langle 101|R^{t}|s\rangle|^2$, as shown in figure \ref{figosc}
($t=0,1,2,\ldots$).
For the TDMPS simulation, we used five qubits including a single oracle qubit and
a single ancilla qubit. The oracle circuit was simply constructed by three CCNOT gates
and two NOT gates. This is just for
simplified demonstration; a meaningful quantum search should be performed for a
realistic computational problem like a satisfiability problem \cite{GJ79} with a
large input size. This will be hopefully realized in future
with a real quantum computer. The quantum circuit of the total process of our
present concern is quite simple as illustrated in figure \ref{figGC}.
In addition, here, truncations of nonzero Schmidt coefficients were not employed.
%
%
%
%

\begin{figure}[tbp]
\begin{center}
\includegraphics[width=95mm]{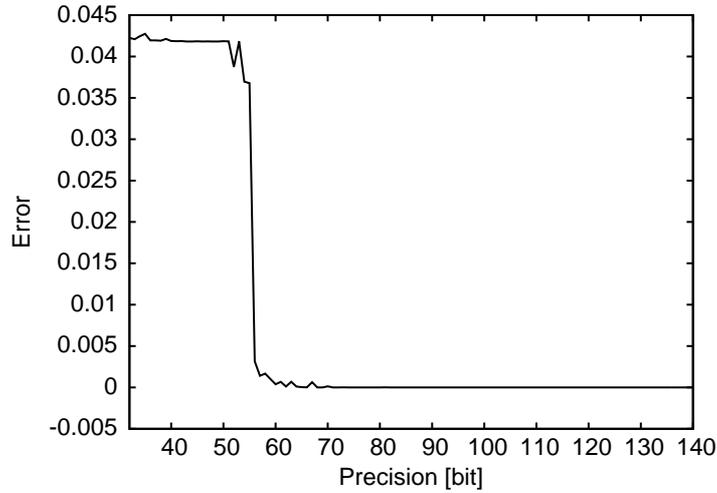}
\caption{\label{fig_prec_error} Computational error
$|\widetilde{\rm Prob}_{101}(20)-{\rm Prob}_{101}(20)|$ 
plotted against the precision.}
\end{center}
\end{figure}

We plot the computational error $|\widetilde{\rm Prob}_{101}(20)-{\rm Prob}_{101}(20)|$ 
after 20 iterations of the Grover routine against the precision
in figure\ \ref{fig_prec_error}, where $\widetilde{\rm Prob}_{101}(20)$ is a
computed value and ${\rm Prob}_{101}(20)=|\langle 101|R^{20}|s\rangle|^2
=\frac{5327874951961}{8796093022208}$
is the exact value calculated symbolically by the Maxima system \cite{Maxima}
with the Qcomp.mac package \cite{Qcompmac}.
\noindent
It is manifest that the double precision (53 bits) is not enough and
more than 70 bits precision is preferable for an accurate simulation.
In addition, it is interesting that the computational error does not
look like a simple elementary function of the precision; there is a sudden drop
at a certain value of the precision.

\begin{figure}[tbp]
\begin{center}
\includegraphics[width=95mm]{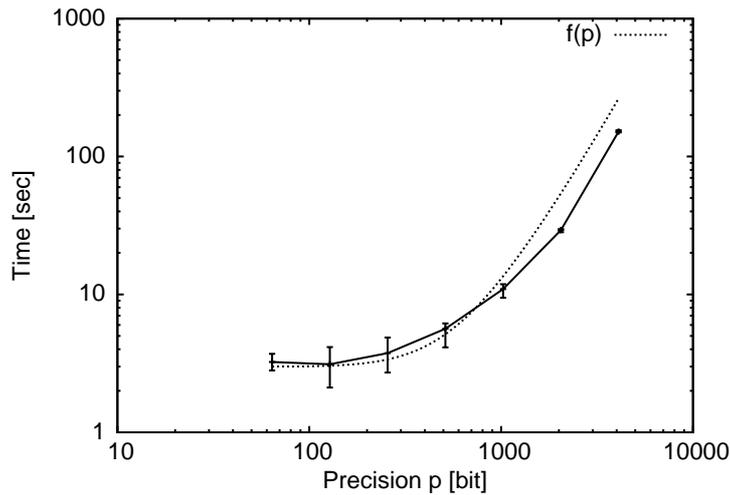}
\caption{\label{fig_prec_time}Computation time to simulate 20 iterations of
the Grover routine, plotted against the precision. 
The average was taken over 10 trials for each data point. Error bars
span from the minimum to the maximum.
Here, $f(p)=1.5\times 10^{-9}(p-64)^3+1.0\times10^{-5}(p-64)^2+3.0$.
Environment: Redhat Enterprise Linux 6 on Intel Xeon X7542 CPU 2.67GHz,
132GB memory.}
\end{center}
\end{figure}

We next evaluate the computational cost by increasing the precision of the TDMPS
simulation. The real CPU time used for simulating the above Grover search
is plotted against the precision in figure \ref{fig_prec_time}.
In the figure, we employed a polynomial for coarse fitting, since
the computational cost of a TDMPS simulation of a quantum algorithm is
guaranteed to be some polynomial in the floating-point precision $p$
(in bits) because each basic arithmetic operation is performed within
polynomial time. It can be exponential in the input size of a problem
instance, but this is not of our present interest. One might expect that
the cost grows quasi-linearly in $p$ considering the cost of each floating-point
multiplication. However, the value of $p$ does effect on the number of
inverse iterations to guarantee enough accuracy in each matrix diagonalization.
In this sense, the fitting seems plausible.
The maximum Schmidt rank of the MPS at the time steps between basic quantum gates
was $4$ during the above simulation. One may find it smaller
than expected. This is because operations $\in {\rm U}(8)$ are simulated
without decomposition in terms of ${\rm U}(4)$ and ${\rm U}(2)$ operations,
in ZKCM\_QC.

\subsubsection{Example 2}\label{secprec2}
\begin{figure}[htbp]
\begin{center}
\includegraphics[width=85mm]{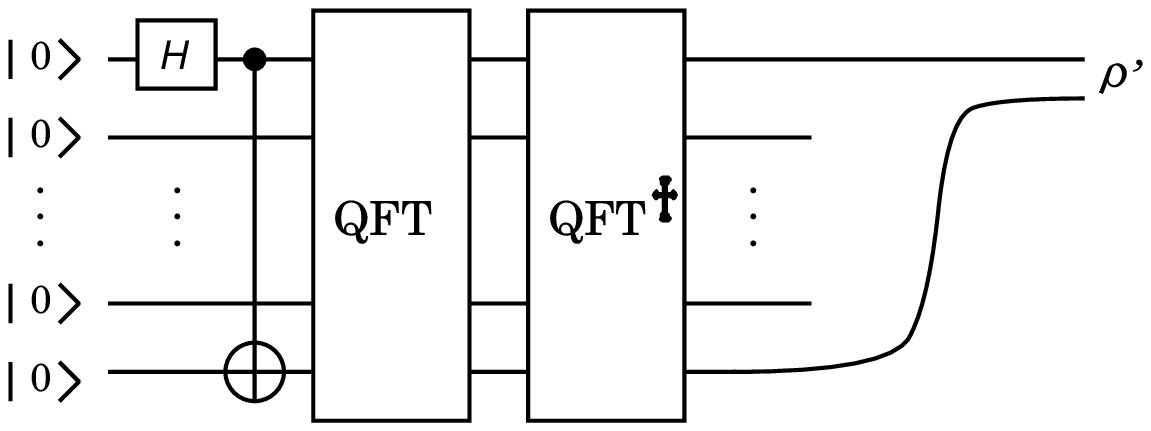}
\caption{\label{fig_fft_circ}Quantum circuit with QFT and its inverse (${\rm QFT}^\dagger$).
This circuit has no practical meaning but to examine a numerical error caused by precision
shortage in a TDMPS simulation.}
\end{center}
\end{figure}

The sudden drop of the numerical error found in figure \ref{fig_prec_error} is
not specific to the example. We will now find a similar behavior in another example.
Here, we consider an $n$-qubit quantum circuit shown in figure \ref{fig_fft_circ}.
A Bell state $|\phi\rangle_{0,n-1}=
(|0\rangle_0|0\rangle_{n-1}+|1\rangle_0|1\rangle_{n-1})/\sqrt{2}$ of qubits $0$ and $n-1$ is
initially created and then the state $|\phi\rangle_{0,n-1}|0\cdots0\rangle_{1,\ldots,n-2}$
goes through the quantum Fourier transform (QFT) and its inverse. Thus, the resultant
state of the qubits $0$ and $n-1$ should be the initial Bell state. Here, we employ
Fowler {\em et al.}'s linear-nearest-neighbour construction for QFT \cite{FDH04} exactly
as it is. 
\begin{figure}[htbp]
\begin{center}
\includegraphics[width=151mm,height=75mm]{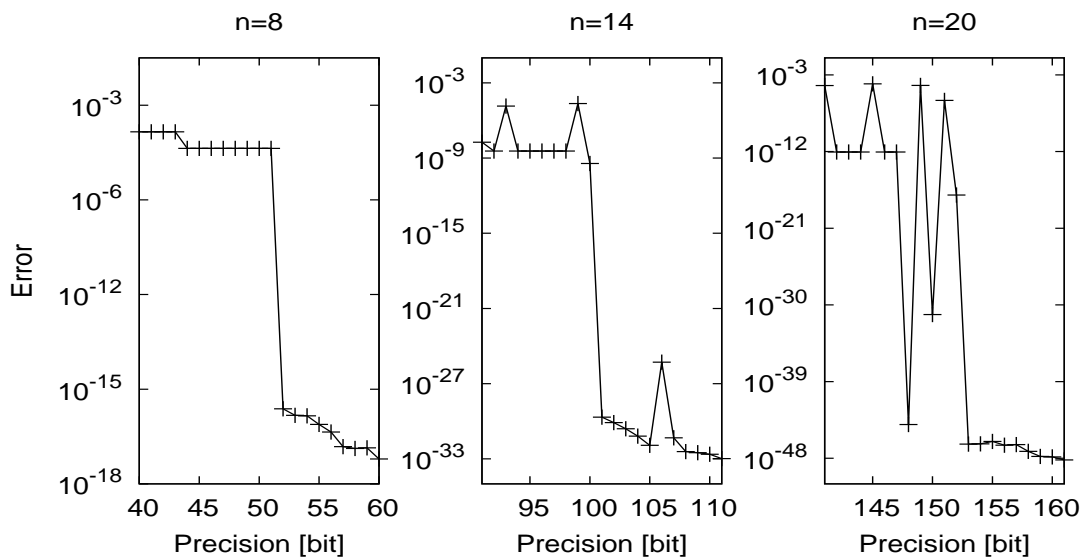}
\caption{\label{fig_qft_prec}Precision dependence of the numerical error
$|\langle00|\widetilde{\rho'}|11\rangle-1/2|$ for the three values of $n$.}
\end{center}
\end{figure}

We performed TDMPS simulations of the circuit for $n=8$, $14$, and $20$ and examined the
precision dependence of the numerical error in the $|00\rangle\langle 11|$ element of
the reduced density matrix $\rho'$ of the qubits $0$ and $n-1$ in the resultant state.\footnote{
It took approximately $1.88$ seconds for each run of the simulation to obtain $\widetilde{\rho'}$
when $n=20$ and precision was set to $150$ bits, in the environment: Redhat Enterprise Linux 6
on Intel Xeon X7542 CPU 2.67GHz, 132GB memory. This fast simulation is not surprising since QFT
maps a computational basis state to a product state (see, {\em e.g.}, reference~\cite{Draper}).}
Truncations of nonzero Schmidt coefficients were not employed.
Ideally, we should have $\langle00|\rho'|11\rangle=1/2$. Let $\widetilde{\rho'}$ be
the computed matrix; the numerical error is quantified as
$|\langle00|\widetilde{\rho'}|11\rangle-1/2|$.
We found that the sudden drop of the error appeared for all of the three values of $n$
as shown in figure \ref{fig_qft_prec} (Note that the vertical axis is in the logarithmic scale).
The figure also indicates that the precision required for avoiding an obvious numerical
error tends to grow as the system size grows.\footnote{This is a rare example which exhibits
such a clear size dependence of the required precision, among many quantum circuits so far as
we tried. It is important to notice the existence of such an example; it indicates that one
may perhaps encounter a {\it hard instance} for which a significantly high precision is
mandatory for a relatively large system size.}

So far we have seen two examples, which have clarified the importance of high-precision
computation for accurate TDMPS simulations of quantum computing. We will next
examine the truncation error of TDMPS.

\subsection{Errors due to truncations}\label{sectrunc}
It is uncommon to make use of a truncation of nonzero Schmidt coefficients in TDMPS simulations
of quantum computing unlike those used for condensed matter physics as we mentioned in
section~\ref{secintro}. We however investigate the case we dare to impose truncations.
Let us denote the limit on the number of Schmidt coefficients that can be
kept for each splitting as $m_{\rm trunc}$. Thus, among Schmidt coefficients,
larger $m_{\rm trunc}$ coefficients are kept and the remaining small coefficients
are truncated out at each step in the simulation of a quantum circuit.

Here we study the effect of a truncation in a TDMPS simulation of the
Deutsch-Jozsa algorithm \cite{DJ92} for a function $f:\{0,1\}^9\rightarrow
\{0,1\}$ with a nine-bit input ${\bf x}$. Recall that $f$ is either
balanced ({\em i.e.}, $\mathit{\sharp}\{{\bf x}|f({\bf x})=0\}
=\mathit{\sharp}\{{\bf x}|f({\bf x})=1\}$) or constant
({\em i.e.}, $f({\bf x})$ is same for all ${\bf x}$) by the promise of
the problem definition for the algorithm (see, {\em e.g.}, section~3.1.2 of \cite{Gruska}).
The Deutsch-Jozsa algorithm for nine qubits is interpreted as the following
process. (i) Apply $H^{\otimes 9}V_fH^{\otimes 9}$ to $|0_0\ldots 0_8\rangle$
(here, $V_f$ is an operation to put the factor $(-1)^{f({\bf x})}$
to the states $|{\bf x}\rangle$); (ii) Measure the nine qubits of the
resultant state. When $f$ is balanced, the probability of having the nine
qubits in $|0\rangle$'s simultaneously in the resultant state vanishes;
when $f$ is constant, the probability is exactly unity.

\begin{figure}[htbp]
\begin{center}
\includegraphics[width=105mm]{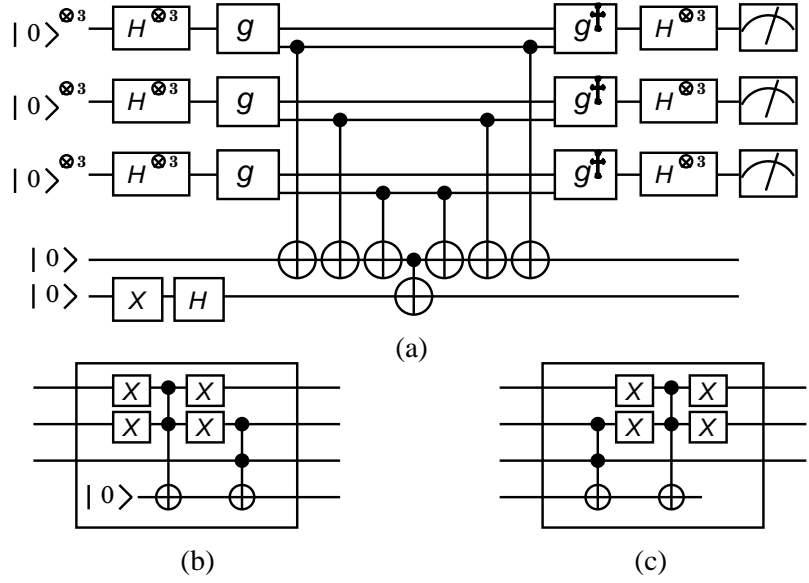}
\caption{\label{figDJf}Quantum circuit of the Deutsch-Jozsa algorithm for
the function $f$. Circuits (b) and (c) describe the gates $g$ and $g^\dagger$ of
circuit (a), respectively.}
\end{center}
\end{figure}

\begin{figure}[tbp]
\begin{center}
\includegraphics[width=95mm]{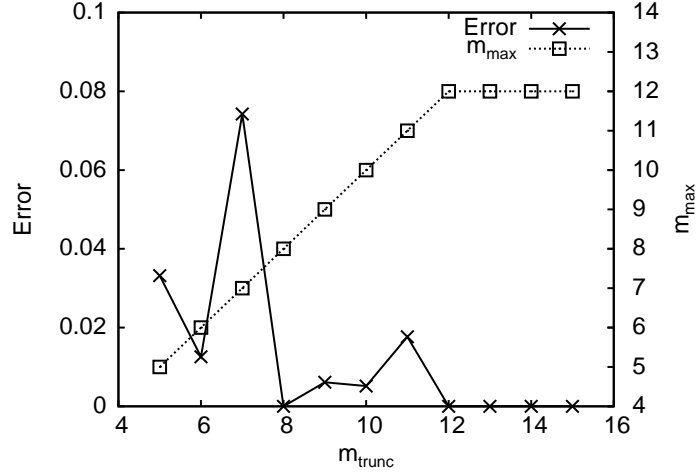}
\caption{\label{fig_trunc_error} Plot of the error in the computed value
of ${\rm Prob}(0_00_10_2)$ (crosses) and the plot of $m_{\rm max}$ (boxes)
as functions of $m_{\rm trunc}$. See the text for the details of the simulation.}
\end{center}
\end{figure}

We employ the following function for our TDMPS simulation.
$f({\bf x})=g(x_0x_1x_2)\oplus g(x_3x_4x_5)\oplus g(x_6x_7x_8)$ with
$g({\bf y})=[(\neg y_0)\wedge(\neg y_1)]\vee (y_1\wedge y_2)$.
This function is balanced. In addition, more specifically, we have
${\rm Prob}(0_00_10_2)=0$ in the resultant state of (i) for this
particular function.
The quantum circuit of the Deutsch-Jozsa algorithm for this function is illustrated
in figure \ref{figDJf}.
%
%
%
%
%
We set the floating-point precision to 256 bits and tried several different values of
$m_{\rm trunc}$ in the TDMPS simulation of the quantum circuit.\footnote{It took less than
$10.2$ seconds for each run of the simulation in the aforementioned environment.}
As shown in figure~\ref{fig_trunc_error},
the error in the computed value of ${\rm Prob}(0_00_10_2)$ (namely, a discrepancy from zero
in the present context) vanishes for $m_{\rm trunc}\ge 12$ while a considerable error
exists for $m_{\rm trunc} \le 11$.
The largest value among $m_s$'s during the simulation is shown as $m_{\rm max}$ in the figure.
It indicates that $m_{\rm trunc}$ should be set to at least the maximum possible value
of $m_{\rm max}$ which is $12$ in the present case, so as to avoid a nonnegligible error.

\begin{figure}[tbp]
\begin{center}
\includegraphics[width=95mm]{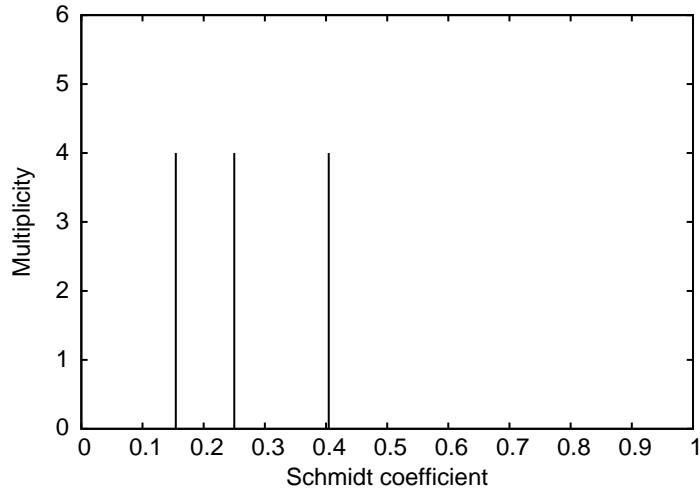}
\caption{\label{figcoeff}Distribution of Schmidt coefficients for the splitting
at the points we had the maximum Schmidt rank in the simulation. The distribution
was same for both of the points.
}
\end{center}
\end{figure}

The above result shows that we cannot truncate out any nonzero Schmidt coefficient during
the simulation. This phenomenon should be related to the distribution of Schmidt
coefficients if discussions of reference \cite{VDMB09} apply to the present case. We show
the distribution at the points we had the maximum Schmidt rank $12$ in the simulation, in figure
\ref{figcoeff} (this is same for both of the points). These points were the second CNOT gate
from the left and one from the right among the seven CNOT gates in the middle part of
circuit (a) of figure \ref{figDJf}. It is now manifest that none of the twelve Schmidt
coefficients is negligible. This is how even a single truncation caused a significant error.

\section{Discussion}\label{secdiscussion}
The TDMPS method is regarded as an approximation method in condensed
matter physics, but is usually an exact method in simulation studies of
quantum computing. Truncations of nonzero small Schmidt coefficients
are thus unemployed, usually, for the purpose of simulating quantum computation.
In a known case in condensed matter physics \cite{VDMB09} where we should avoid
truncations, a dominant number of Schmidt coefficients are nonnegligible. This
should be true also in TDMPS simulations of quantum circuits.
In fact, we found that this is the case and the threshold $m_{\rm trunc}$ for
the largest number of surviving Schmidt coefficients should be at least the maximum
possible number of nonzero Schmidt coefficients, in the simulation results for the
Deutsch-Jozsa algorithm presented in section\ \ref{sectrunc}. 

Besides, we have also evaluated the effect of precision shortage. Rounding errors of
individual basic arithmetic operations may accumulate to a large error in the
resultant state of a TDMPS simulation of a quantum circuit.
In our simulation of a simple quantum search presented in section\ \ref{secprec1},
a nonnegligible error was observed in the resultant population of a target state
unless we increased the precision beyond the double precision.
Furthermore, we found that the error does not behave like a smooth curve as a function
of the precision in bits, but behaves like a step function with a sudden drop, as
shown in figure\ \ref{fig_prec_error}.
This phenomenon was also observed
in the TDMPS simulation of a quantum circuit containing the quantum Fourier transform,
for three different system sizes, as presented in section\ \ref{secprec2}. In addition, 
this is not a particular phenomenon for a TDMPS simulation but a rather commonly
observed one for matrix computation (see the document of the ZKCM library \cite{ZKCM}).

With our simulation results for investigating numerical errors, it is suggested that
the precision should be preferably at least 70 bits and the truncation of nonzero Schmidt
coefficients is not encouraged for a TDMPS simulation of quantum computation.
As for the cost of multiprecision computation, figure\ \ref{fig_prec_time} indicates that
more than 1000 bits precision is quite expensive. As a matter of fact, the current computer
architecture does not have a good hardware support for high precision computation. Thus
the practical precision we may employ should be several hundreds bits for the time being.

\section{Conclusion}\label{secconclusion}
We have utilized our multiprecision TDMPS library ZKCM\_QC to investigate
numerical errors in TDMPS simulations of quantum computing.
To avoid a nonnegligible rounding error within a practical cost, it is suggested
that the length of the mantissa portion of each floating-point number should be
beyond the double-precision length but not more than 1000 bits in the present
technology. It is also suggested that a truncation of nonzero Schmidt coefficients
is discouraged.

\section{Software information}
The simulations were performed with ZKCM\_QC ver. 0.0.9beta put on the
repository\\ \verb|https://sourceforge.net/p/zkcm/sublibqc|~.

\section*{References}
\bibliographystyle{iopart-num}
\bibliography{refs_ccp}

\end{document}